\documentclass[usenatbib]{mnras}

\usepackage{natbib}
\bibpunct{(}{)}{;}{a}{}{,}
\usepackage{graphicx}
\usepackage{txfonts}
\usepackage{xcolor}
\usepackage{multirow}
\usepackage{pst-plot}
\usepackage{breakurl}
\SpecialCoor
\usepackage{hyperref}

\def\inte{{\em INTEGRAL}}
\def\xmm{{\em XMM-Newton}}

\def\chan{{\em Chandra}}
\def\asca{{\em ASCA}}
\def\beppo{{\em BeppoSAX}}
\def\rxte{{\em RXTE}}
\def\swift{{\em Swift}}

\def\suzaku{{\em Suzaku}}
\def\nustar{{\em NuSTAR}}

\def\maxi{{\em MAXI}}

\def\ferg{\mathrm{erg\,s^{-1}\,cm^{-2}}}

\def \src {AX\,J1841.0-0536}
\def \srcb {SAX\,J1818.6-1703}

\definecolor{arancio}{rgb}{1,0.5,0}
\definecolor{viola}{rgb}{0.7,0,1}
\definecolor{verde}{rgb}{0.2,0.7,0.7}
\definecolor{cobalt}{rgb}{0.0, 0.28, 0.67}
\definecolor{airforceblue}{rgb}{0.36, 0.54, 0.66}
\definecolor{ballblue}{rgb}{0.13, 0.67, 0.8}
\definecolor{battleshipgrey}{rgb}{0.52, 0.52, 0.51}
\definecolor{darkgreen}{rgb}{0.0, 0.2, 0.13}

\title[X-ray observations of AX\,J1841.0-0536 and SAX\,J1818.6-1703]{\nustar\ and \swift\ observations of two supergiant fast X-ray transients: AX\,J1841.0-0536 and SAX\,J1818.6-1703}
\author[E. Bozzo et al.]{
E.\ Bozzo,$^{1}$\thanks{E-mail: enrico.bozzo@unige.ch}
C.\ Ferrigno,$^{1}$
P.\ Romano,$^{2}$ \\
$^{1}$Department of Astronomy, University of Geneva, Chemin d'Ecogia 16, CH-1290 Versoix, Switzerland\\
$^{2}$INAF, Osservatorio Astronomico di Brera, Via E.\ Bianchi 46, I-23807, Merate, Italy
}

\date{}

\begin{document}

\maketitle

\begin{abstract}
Supergiant fast X-ray transients are wind-fed binaries hosting neutron star accretors, which display a peculiar variability in the X-ray domain. Different models have been proposed to explain this variability and the strength of the compact object magnetic field is generally considered a key parameter to discriminate among possible scenarios. We present here the analysis of two simultaneous observational campaigns carried out with \swift\ and \nustar\ targeting the supergiant fast X-ray transient sources \src\ and \srcb.\ A detailed spectral analysis is presented for both sources, with the main goal of hunting for cyclotron resonant scattering features that can provide a direct measurement of the neutron star magnetic field intensity. \src\ was caught during the observational campaign at a relatively low flux. The source broad-band spectrum was featureless and could be well described by using a combination of a hot blackbody and a power-law component with no measurable cut-off energy. In the case of \srcb,\ the broad-band spectrum presented a relatively complex curvature which could be described by an absorbed cut-off power-law (including both a cut-off and a folding energy) and featured a prominent edge at $\sim$7~keV, compatible with being associated to the presence of a ``screen'' of neutral material partly obscuring the X-ray source. The fit to the broad-band spectrum also required the addition of a moderately broad ($\sim$1.6~keV) feature centered at $\sim$14~keV. If interpreted as a cyclotron resonant scattering feature, our results would indicate for \srcb\ a relatively low magnetized neutron star ($\sim$1.2$\times$10$^{12}$~G).
\end{abstract}


\section{Introduction}
\label{sec:intro}

Supergiant fast X-ray transients (SFXTs) are a peculiar sub-class of the long-known classical supergiant X-ray binaries (SgXBs), hosting a neutron star (NS) accreting from the wind of its supergiant OB companion. Compared to the classical systems, SFXTs display a much more prominent X-ray variability featuring sporadic hour-long outbursts that can reach a luminosity up to a few times $\sim$10$^{38}$ erg s$^{-1}$ \citep[see, e.g.][for a recent overview]{romano23}. In between the outbursts, SFXTs spend long intervals of time in a lower luminosity state ($\sim$10$^{33-34}$~erg s$^{-1}$) during which frequent fainter X-ray flares take place. These have durations similar to that of the brightest outbursts and are characterized by similar spectral variations \citep[see, e.g.,][for recent reviews]{nunez17,kre20}. The lowest luminosity state of the SFXTs, usually termed ``quiescence'', can be as faint as 10$^{31-32}$~erg s$^{-1}$, often out of reach even for the most sensitive X-ray operating facilities \citep{bozzo09,bozzo10,sidoli10,bodaghee11,sidoli23}. The typical dynamic range of the X-ray luminosity in SFXTs is of 10$^5$, with the record value achieved so far only by IGR\,J17544-2619 and extending up to $\sim$10$^{6}$ \citep[][]{romano15b}. Long term studies of the X-ray emission from SFXTs in both the hard \citep{paizis14} and soft X-ray domain \citep{romano14} have demonstrated that SFXTs are dramatically sub-luminous compared to classical SgXBs, with a typical reduction in the expected X-ray luminosity up to a factor of 10$^3$-10$^{4}$ \citep{bozzo15}.

Although almost 20~years have passed since the first recognition of the SFXTs as a class \citep{sguera06}, the reason beyond their extreme variability is still actively debated. The two most widely discussed possibilities are that either magnetic/centrifugal gatings \citep[associated with the NS spin and magnetic field strength;][]{bozzo08} or a settling accretion regime combined with somewhat less powerful stellar winds than in  classical systems \citep{shakura12} are at work to substantially reduce the average mass accretion rate onto the compact object in SFXTs. The former mechanism would be favoured if the NSs in the SFXTs are endowed with high magnetic fields ($\gtrsim$10$^{13}$-10$^{14}$~G) and long spin periods ($\gtrsim$100-1000~s), while the latter model to be operational at the needed efficiency requires the wind of the OB supergiants in the SFXTs to be magnetized and, on average, somewhat less powerful (in terms of density and velocity) compared to the stellar winds in classical systems. The long-term inhibition of accretion in both the above models can explain the fact that the SFXTs are largely sub-luminous compared to the classical systems, but both models require the presence of ``clumps'' in the wind of the supergiant companion to trigger the SFXT flares/outburst and achieve the required dynamic range in the X-ray luminosity. Clumps are inhomogeneities in the winds of OB supergiants and are known to be substantially over-dense compared to the intra-clump medium \citep[see, e.g.,][for a recent review]{rubke23}.
In the gating model, the encounter between the NS and a dense clump can sufficiently compress the magnetosphere of the compact object to temporarily restore a high mass accretion rate and lead to large short-lived X-ray emission which is comparable to an SFXT flare or outburst \citep[depending mainly on the gating mechanism at work and the size of the clump;][]{bozzo08,bozzo17,ferrigno22}. In the settling accretion model, the flare/outburst due to the encounter between the NS and a clump can only occur if the clump has dragged with it the magnetic field of the supergiant, leading to reconnections with the NS magnetosphere. This causes the collapse of a hot shell surrounding the compact object onto the NS, leading to a temporary high mass accretion rate \citep[and thus high X-ray luminosity of the source; see][]{shakura14}. Discriminating among the different proposed models to explain the SFXT phenomenology thus requires an in-depth study of the NS spin and magnetic field on one side, as well as the stellar wind properties on the other (including, but not limited to, the clump properties).

Over the years, a large body of evidence has been collected on the presence of clumps around the compact objects in SFXTs. This is either ``direct'' observational evidence, when the clump is accreted by the NS and causes highly noticeable variations in the X-ray flux and in the local absorption column density toward the source \citep{bozzo11}, or ``indirect'' evidence, when the clump is simply passing in front of the compact object causing a temporary (partial) obscuration of the X-ray source \citep[see, e.g.][]{Walter2007,rampy09,bozzo17,ferrigno22}.
Measurements of the stellar wind properties in SFXTs proved very challenging due to their distance (usually up to several kpc) and the intrinsic absorption. So far, there has been  tentative evidence for the stellar wind in one SFXT to be possibly magnetized \citep{hubrig18} but no quantitative evidence in favor of a systematic difference between the stellar winds in SFXTs and those in classical systems \citep[see][and references therein]{pragati18,hainich20}. Measurements of the NS magnetic fields and spin periods in SFXTs remain so far elusive, with practically no secured results for all confirmed sources in this class. Concerning the spin periods, the only measurement available to date for a confirmed SFXT source is at $\sim$187~s and pertains the sole periodically outbursting SFXT IGR\,J11215–5952 \citep{Sidoli2006,sidoli13,bozzo15,romano15,bhalerao15}. A spin period of $\sim$1212~s has been also reported multiple times in the case of the source IGR\,J16418-4532 \citep{sidoli12,drave13}, but the belonging of this source to the SFXT class is still debated due to the relatively limited dynamic range displayed in the X-ray domain \citep{romano12f}. NS magnetic field measurements have been attempted for several objects especially in the most recent years, exploiting the unique combination of sensitivity and broad-band coverage of the instruments on-board \nustar.\ Different authors have looked in the X-ray spectra of the SFXTs for the presence of cyclotron scattering resonant features \citep[CRSFs; see][for a recent review]{staubert19}. These features are known to provide the most direct evidence and robust estimate of NS magnetic fields in the range of a few 10$^{12}$~G within an expected accuracy of about 30\% \citep[see, e.g.][]{poutanen13,mushtukov15}. A detection was reported in the case of the SFXT prototype IGR\,J17544-2619 indicating a relatively low magnetic field of $\sim$10$^{12}$~G \citep{bhalerao15}, but a later observation with a higher signal-to-noise ratio (S/N) did not confirm such detection \citep{bozzo17544}. Tentative detections of CRSFs were also reported in the cases of the SFXT IGR\,J18483-0311 and IGR\,J11215-5952 \citep[see][respectively]{sguera10,sidoli17} but again never confirmed by deeper observations \citep{ducci13b,sidoli20}.

In this paper, we pursue the hunt for CRSFs in the X-ray spectra of SFXTs by analyzing two simultaneous sets of \swift\ and \nustar\ observations carried out in the direction of the confirmed SFXT sources \src,\ and \srcb.\ We briefly summarize the present knowledge on these systems in Sect.~\ref{sec:sources} and provide all relevant information concerning the exploited observations in Sect.~\ref{sec:data}. All results are presented in Sect.~\ref{sec:results}. We finally provide our conclusions in Sect.~\ref{sec:conclusions}. All uncertainties in this paper are given at 90\% confidence limit, unless stated differently.

\section{The source sample}
\label{sec:sources}

\subsection{\src}
\label{sec:src}

\src\ was discovered by \asca\ in 1994 during a moderate flaring activity \citep{bamba01} and subsequently observed in quiescence by \chan,\ whose fine localization capabilities led to the identification of the companion star as a B1\,Ib supergiant \citep{halpern04,nespoli08}. The distance to the source remains so far rather uncertain, but most likely $\gtrsim$10~kpc \citep{gaia21,fortin22}. The source has been detected in outburst many times in the past, triggering the instruments on-board \inte,\ \maxi,\ and \swift\ \citep{rodriguez04,sidoli08,romano09,romano10,depasquale10,negoro10,romano12,romano12c,romano23}. In 2011, a bright flare was caught serendipitously by \xmm,\ leading to possibly the most striking evidence for the presence of stellar wind clumps in SFXTs \citep{bozzo11}. Fainter flares have been detected in high sensitivity observations carried out with both \xmm\ and \suzaku,\ often revealing prominent spectral variability. In most cases, this was driven by sudden changes in the local absorption column density and ascribed to the presence of dense clumps in the environment surrounding the NS \citep{2012PASJ...64...99N,bozzo17}. The system orbital period remains so far elusive, while the tentative detection of a 4.7~s spin period by \asca\ \citep[][]{bamba01} has been questioned by \citet{bozzo11} and never confirmed in other data-sets. The source was also targeted during an exploratory observational campaign in the mm band, but remained undetected \citep{mmband}.

The broad-band X-ray spectrum of \src\ has been investigated by using both \suzaku\ and \swift\ observations. In the former case, \citet{2012PASJ...64...99N} exploited the combination of the XIS and HXD/PIN instruments to obtain a 0.1-40~keV spectrum that could be satisfactorily fit with a model comprising a partial covering component, a cut-off power-law, and a iron emission line centered at 6.4~keV. The authors also included in the fit an absorption-like feature centered at $\sim$15~keV and reminiscent of a CRSF. No details were provided concerning the statistical significance of such a feature. The combined \swift/XRT and BAT spectra (0.3-100~keV) presented by \citet{romano12} could be  described well by using a simple cut-off power-law model, with no evidence for the presence of the iron line, the partial covering, and the tentative CRSF.

\subsection{\srcb}
\label{sec:srcb}

\srcb\ was discovered by \beppo\ during an outburst in 1998 \citep{zand98}.
Since then, the source has been triggering the instruments on-board both \swift\ and \inte\ many times,
following the onset of the typical SFXT outbursts \citep[see, e.g.,][and references therein]{bird09,bird09b,paizis14,romano14b,romano23}.
Follow-up observations of the source in quiescence with \chan\ allowed \citet{torrejon10} to confirm the
nature of the donor star in this system to be a B0.5\,Iab supergiant at a distance of 2.1~kpc \citep[see also][]{negueruela06}.
There is yet no indication about pulsations from the source, but the orbital period has been firmly established at 30~days and the associated eccentricity has been estimated at 0.3-0.4 \citep{bird09b,zurita09}. By exploiting all archival \inte\ data, \citet{bird09b} verified that most of the source outbursts (about 60\%) occur close to periastron, while the deepest upper limits on the source non detections derived from \xmm\ data \citep{bozzo09,bozzo12} corresponded to orbital phases close to the apastron.

In the soft X-ray domain, \srcb\ has displayed a frequent remarkable spectral variability which has been often ascribed to changes in the absorption column density due to the clumpy wind accretion. The average absorption column density local to the source is among the highest recorded in SFXTs, peaking at $\gtrsim$10$^{23}$~cm$^{-2}$ \citep{boon16,bozzo17}. The broad-band X-ray spectrum of the source remained so far poorly studied. In the combined \xmm\ and \inte\ dataset presented by \citet{boon16}, the \inte\ data were characterized by a very limited statistics and could not help constraining the spectral shape above $\sim$10~keV. The combined XRT+BAT spectrum collected during an outburst of the source caught by \swift\ in 2009 could be described well by using either a simple absorbed cut-off power-law or a comptonization model, with no indication for the presence of emission and absorption features \citep{sidoli09}.
\begin{figure*}
  \centering
  \includegraphics[width=9.0cm,angle=-90]{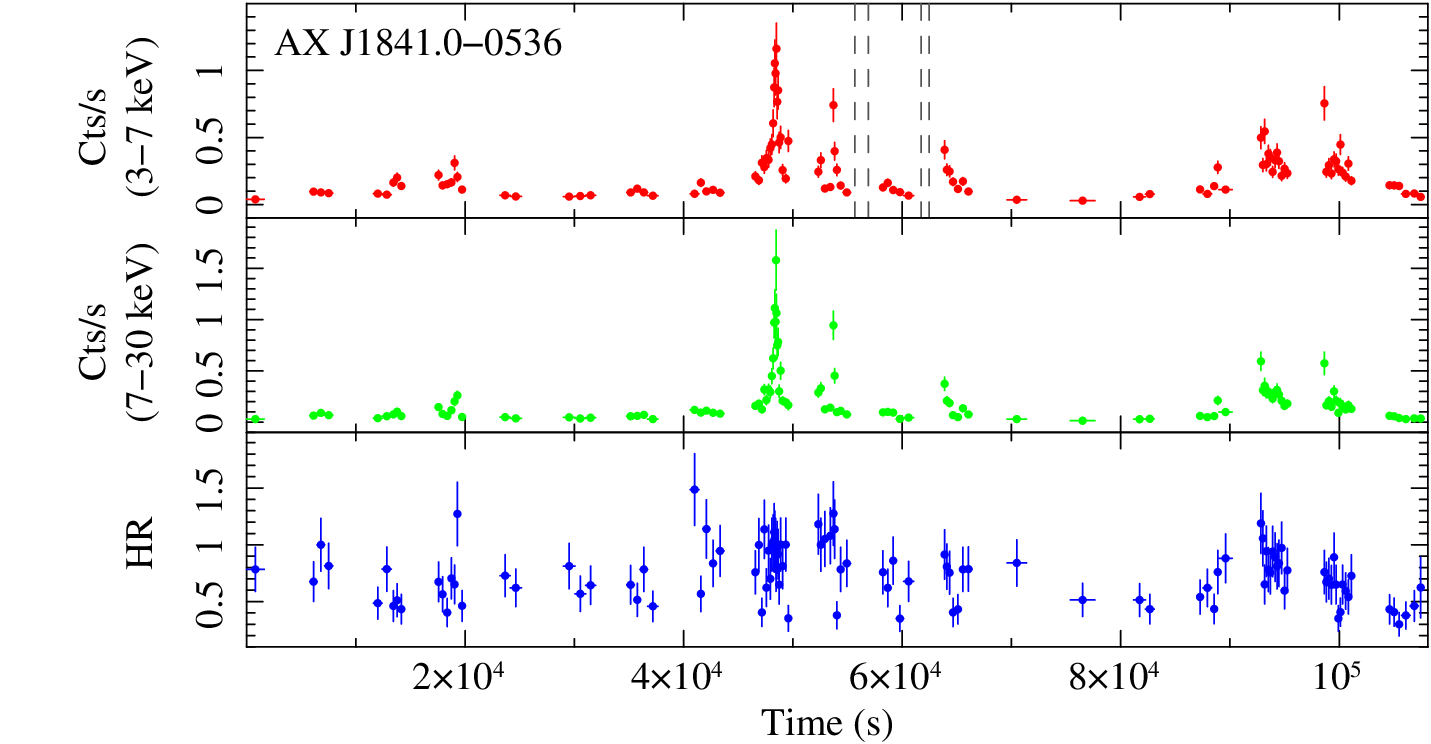}
  \caption{\label{fig:nustar_lc1} Adaptively rebinned energy-resolved lightcurves of \src\ as observed by \nustar/FPMA during the ObsID~30301026002 (the reference time for the x-axis is 58041.9864\,MJD). The upper panel shows the lightcurve in the 3--7~keV energy range, while the mid-panel shows the lightcurve in the 7--30~keV energy range. The bottom panel shows the adaptively rebinned HR. The dashed vertical lines mark the intervals over which XRT data were collected. We marked explicitly the timing boundaries of the different snapshots included in the \swift\ observation, as to highlight the effective time spent on the source by the XRT instrument.}
\end{figure*}
\begin{figure*}
  \centering
  \includegraphics[width=9.0cm,angle=-90]{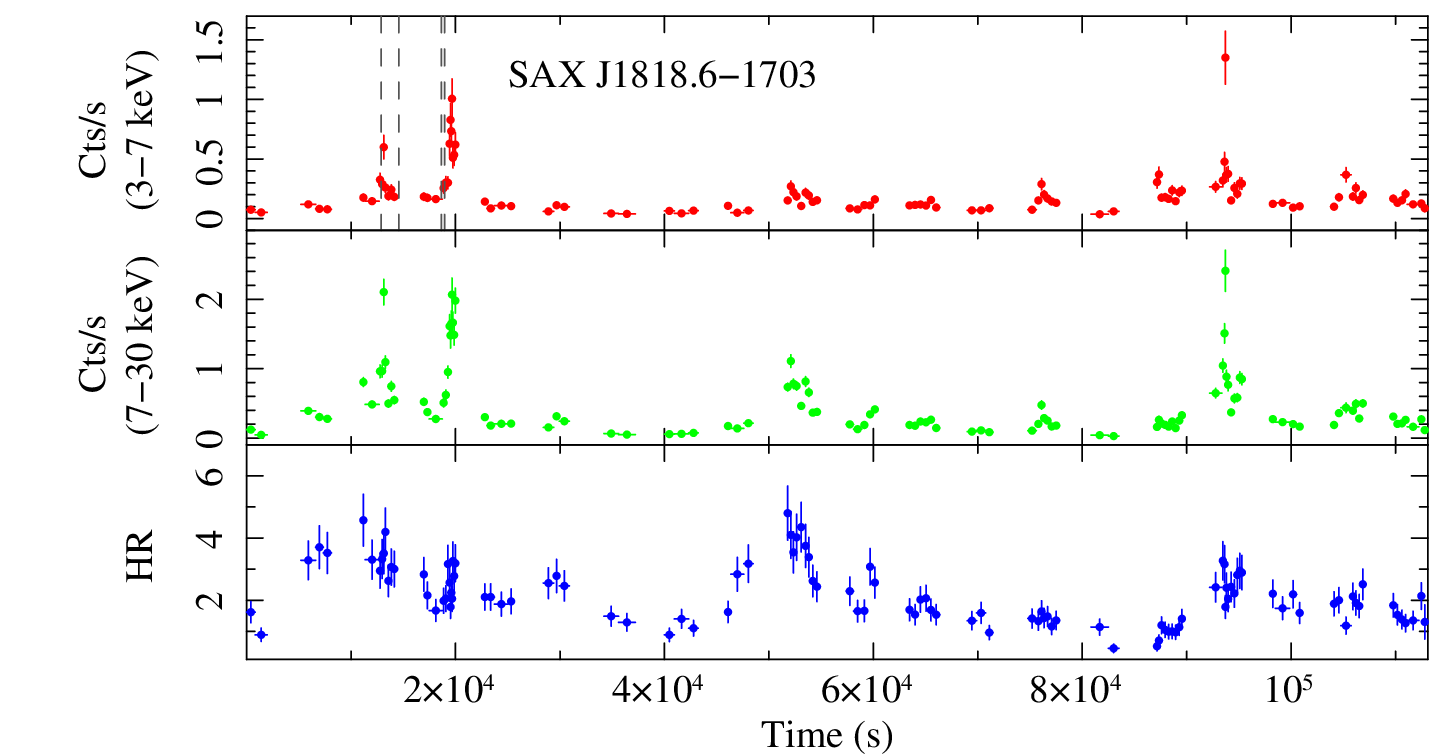}
\caption{\label{fig:nustar_lc2} Same as Fig.~\ref{fig:nustar_lc1} but for the case of \srcb\ (ObsID~30301025002). The reference time for the x-axis is 58171.1923\,MJD.}
\end{figure*}

\section{Data processing and analysis}
\label{sec:data}

\subsection{\nustar}
\label{sec:nustar}

\src\ was observed by \nustar\ on 2017 October 15 for a total exposure time of 54.3~ks (ObsID~30301026002).
We processed the data using standard methods and procedures\footnote{\url{https://heasarc.gsfc.nasa.gov/docs/nustar/analysis/nustar_swguide.pdf}}, exploiting the {\sc nupipeline} v.\,0.4.9 distributed with the {\sc Heasoft} software v.\,6.32.1.
The latest calibration files available at the moment of writing were used for the data reduction (caldb v.\,20230918).
No interval of high background emission and/or evidence for straylight issues were found, so the entire exposure time of the
observation was available for the scientific analysis. The source photons were extracted from a 60\arcsec circle centred on the source,
while the background was evaluated using a similar circular region located in a chip not contaminated from the source emission. We verified that reasonably different choices for both the source and background photons extraction regions did not affect the presented results of the scientific analysis. Each of the \nustar\ FPMA and FPMB recorded during the observation an average source count-rate of 0.13$\pm$0.02~cts~s$^{-1}$ in the 2-40~keV energy band. 

The FPMA and FPMB energy-resolved lightcurves were extracted using the {\sc nuproduct} tool in the 3--7~keV and 7--30~keV bands, computing the corresponding hardness-ratio (HR) via our  adaptive rebinning algorithm \citep{ferrigno22}. The lightcurves were rebinned by the algorithm to reach a S/N of at least 6 in the soft band curve and then the same rebinning is applied to the hard band curve before computing the HR. The result is reported in Fig.~\ref{fig:nustar_lc1} (only showing the FPMA lightcurves as an example). The source remained relatively faint for most of the observation, but displayed a modest flare about 20~ks after the beginning of the exposure. We could not detect any significant variability of the HR across the observation and during the rise/decay of the flare. Therefore, a single spectrum has been extracted using the entire exposure time of the observation. All spectra in this paper were rebinned using the algorithm developed by \citet{Kaastra} and implemented within {\sc Heasoft} via the {\sc ftgrouppha} tool. C-statistics has been used to evaluate the fits \citep{cash79}. We discuss the spectral analysis in Sect.~\ref{sec:results}.

We also extracted the source and background event files from both the FPMA and FPMB, applying a barycentric correction with the tool {\sc barycorr}. From these event files we extracted a light curve binned at 0.1\,s for both units in the band 3--30\,keV. These were summed and inspected for the presence of periodicity in the range 3$\times10^{-5}$-5.45\,Hz through the  Lomb-Scargle periodogram with oversampling of factor 2. The periodogram of the source X-ray emission displayed a modest red noise at frequencies up to 2--3\,mHz. Above this frequency, no statistically significant ($\gtrsim$5$\sigma$) peak is observed.

The \nustar\ observation of \srcb\ was carried out on 2018 February 22 for a total exposure time of 54.5~ks (ObsID~30301025002).
We applied the same data reduction techniques as for the other source. Also in the case of the ObsID~30301025002, no high background time intervals were found and thus we retained the entire exposure time available for the scientific analysis. Some straylight was observed in a corner of the FPMA detector far from the source location, and thus the source and background extraction regions were chosen as to avoid this contamination. We used a source (background) extraction region with a diameter of 69\arcsec (153\arcsec). We verified that reasonably different choices for both the source and background photons extraction regions did not affect the presented results of the scientific analysis. Each of the \nustar\ FPMA and FPMB recorded during the observation an average source count-rate of 0.24$\pm$0.02~cts~s$^{-1}$ in the 2-40~keV energy band. The energy-resolved lightcurves of the source are shown in Fig.~\ref{fig:nustar_lc2}. The source shows a modest flaring activity, undergoing three relatively short episodes of enhanced X-ray emission (lasting a few ks) roughly spaced by 40~ks during the \nustar\ observation. The HR shows a variability that matches relatively well the behavior of the source count-rate, becoming harder during the enhanced X-ray activity periods. Driven by this finding, we extracted for the FPMA and FPMB in ObsID~30301025002 also two sets of HR-resolved spectra, selecting the time intervals intervals when HR$>$3 and HR$<$2. We discuss all spectra in Sect.~\ref{sec:results}.

As for the case of \src,\ we extracted for \srcb\ the source and background event files from the FPMA, applying a barycentric correction with the tool {\sc barycorr}. We inspected the total light curve binned at 0.1\, and extracted in the energy range 3--30\,keV for the presence of possible periodicities in the range $3\times10^{-5}$-5.11\,Hz through the  Lomb-Scargle periodogram. The periodogram of the source X-ray emission displayed some red noise at frequencies up to 20\,mHz but no significant peak was observed above that frequency.

\subsection{\swift\ }
\label{sec:swift}

Two observations of 2~ks each were performed with the narrow field instrument, the X--ray Telescope \citep[XRT,][]{burrows05} on-board the Neil Gehrels {\it Swift} Observatory \citep[][]{Gehrels2004} simultaneously with the \nustar\ ObsID~30301026002 and 30301025002, respectively, extending the energy coverage of the data-sets down to the softer X-rays. XRT observed \src\ starting from 2017 October 16 at 15:08 until 17:01 (UTC). The XRT observation of \srcb\ was carried out from 2018 February 22 at 08:11 to 09:52 (UTC; see Fig.~\ref{fig:nustar_lc1} and \ref{fig:nustar_lc2}). The \swift\ data 
were processed and analysed using the standard software ({\sc FTOOLS}\footnote{\href{https://heasarc.gsfc.nasa.gov/ftools/ftools_menu.html}{https://heasarc.gsfc.nasa.gov/ftools/ftools\_menu.html.}} v6.31), calibration (CALDB\footnote{\href{https://heasarc.gsfc.nasa.gov/docs/heasarc/caldb/caldb_intro.html}{https://heasarc.gsfc.nasa.gov/docs/heasarc/caldb/caldb\_intro.html.}}  20230725), and methods. Filters to the XRT data were applied via the {\sc xrtpipeline} task (v0.13.7). As both \src\ and \srcb\ were relatively faint during the XRT observations, a single spectrum per source was extracted in each of the XRT pointings. The fit to these spectra is discussed in Sect.~\ref{sec:results}, but we anticipate that the relatively high absorption column density affecting the two sources and the low flux limited the S/N achieved by the XRT observations, with virtually no signal below $\sim$2~keV. We inspected the XRT lightcurves of both sources to look for interesting variability patterns, but the low count-rate and the usual fragmentation of the data associated to the observational constraints of \swift\ hampered any meaningful discrimination between lower and higher emission intervals.

\section{Results}
\label{sec:results}

In order to determine the best model to describe the broad-band continuum of \src\ and \srcb,\ we performed a combined fit of the simultaneous \swift\ and \nustar\ observations for the two sources. The fits are discussed separately for the two objects in the following sub-sections.

\subsection{\src}
\label{sec:src_spe}

\begin{figure}
  \hspace{-0.5cm}
  \includegraphics[width=6.5cm,height=8.9cm,angle=-90]{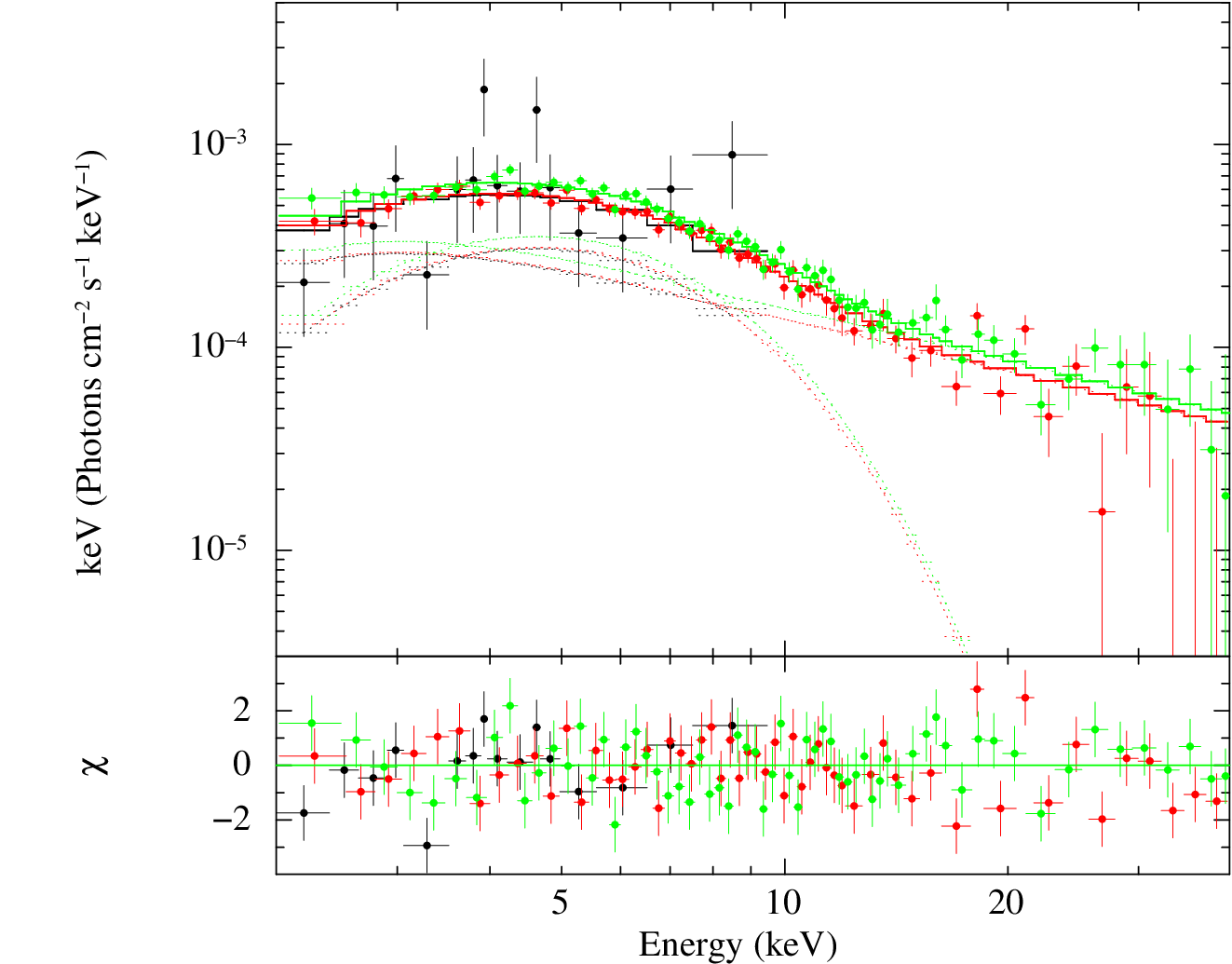}
  \caption{\label{fig:bband1} Unfolded broad-band X-ray spectrum of \src\ as observed by \swift/XRT (black), \nustar/FPMA (red) and \nustar/FPMB (green). The best fit model is an absorbed hot blackbody plus a power law (see Sect.~\ref{sec:src_spe} for details). The residuals from the fit are shown in the bottom panel. For plotting purposes, energy bins have been grouped with the {\sc setplot rebin} within {\sc xspec} to achieve a minimum S/N in each bin of at least 5~$\sigma$.}
\end{figure}

The XRT, FPMA, and FPMB spectra of \src\ were fit together by using the {\sc xspec} environment \citep[v.12.13.1;][]{xspec}. A fit to the three spectra together with a single cut-off power-law component resulted in a reasonable, but not acceptable, description of the data (C-stat/d.o.f.=348.1/291), but wavy residuals from the fit were visible all along the energy coverage provided by the data. In particular, we noticed prominent residuals around $\sim$15~keV that could have been reminiscent of the CRSF feature previously identified in a \suzaku\ observation of the source carried out in 2011 (see Sect.~\ref{sec:src}). The addition of a {\sc gabs} component in {\sc xspec}, centered at 14.2~keV and mimicking the presence of a CRSF, led to a substantial improvement in the shape of the residuals. However, the feature turned out to be unrealistically broad ($\sim$8~keV) and deep ($\sim$37~keV), and thus we did not consider this a plausible description of the broad-band spectrum from \src\ based on the typical properties of CRSFs observed in X-ray pulsars \citep[these features have typical width and depth of 2-3~keV; see, e.g.,][and references therein]{staubert19}. We attempted to improve the fit by adding first a partial covering model ({\sc pcfabs} in {\sc xspec}) and then testing the applicability of different comptonization models, including the {\sc bmc} \citep{bmc} and {\sc compTB} \citep{farinelli08} in {\sc xspec}. None of these possibilities provided a noticeable improvement in the shape of the residuals.
We considered that the addition of a thermal component at the lower energies could have improved the fit and adopted a simple phenomenological model comprising a {\sc bbodyrad} component and a power-law. This model could provide a fully acceptable description of the data (C-stat/d.o.f.=295.2/290), flattening out all significant fluctuations in the residuals from the fit (see Fig.~\ref{fig:bband1}). We measured an absorption column density of $N_{\rm H}$=(2.1$\pm$1.3)$\times$10$^{22}$~cm$^{-2}$ \citep[obtained by exploiting for the fit the {\sc Tbabs} component with the default {\em wilm} abundances and cross sections;][]{vern96,wilms00}, a blackbody temperature and radius of 1.48$\pm$0.06~keV and 0.27$\pm$0.03~km (assuming a distance of 10~kpc, see Sect.~\ref{sec:srcb}), and a power-law photon index of $\Gamma$=1.9$\pm$0.3. The estimated flux in the 2-40~keV energy band was (1.0$^{+0.1}_{-0.3}$)$\times$10$^{-11}$~$\ferg$. We included in the fit normalization constants to take into account inter-calibration uncertainties between the instruments and the fact that the XRT and \nustar\ data did not completely overlap in time (see the top panel of Fig.~\ref{fig:nustar_lc1}). The FPMA normalization was fixed to unity, while the FPMB and XRT ones were measured at 1.08$\pm$0.03 and 1.1$\pm$0.2, respectively.

\subsection{\srcb}
\label{sec:srcb_spe}

\begin{figure*}
\hspace{-1cm}
  \includegraphics[width=12cm,angle=-90]{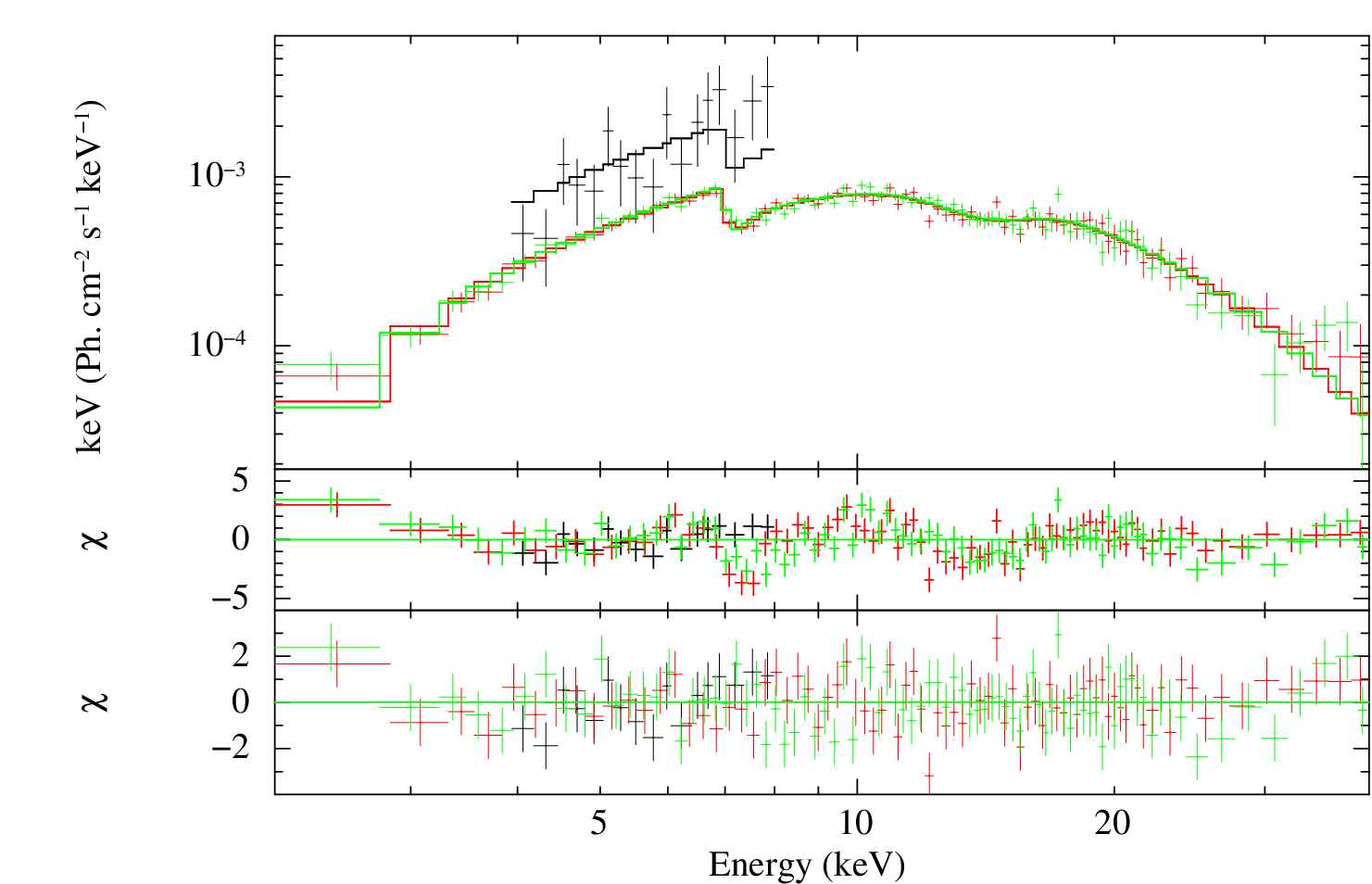}
  \caption{\label{fig:bband2} Same as Fig.~\ref{fig:bband1} but in the case of the source \srcb.\ The best fit model in this case comprises an absorbed cut-off power-law ({\sc Tbabs*highecut*powerlaw} in {\sc xspec}), an absorption feature ({\sc gabs} in {\sc xspec}), and an {\sc edge} centered at $\sim$7~keV (see Sect.~\ref{sec:srcb_spe} for more details). The components of the best fit model are shown in the upper panel of the figure with dashed and dotted lines, while the residuals from the fit are reported in the bottom panel. The mid panel shows the residuals obtained when the broad-band spectra are fit with a simple cut-off power-law.}
\end{figure*}

The XRT, FPMA, and FPMB spectra of \srcb\ extracted by using the entire exposure time available within the \swift\ and \nustar\ observations were fit together first by using a simple absorbed cut-off power-law model. This resulted in a very poor description of the data with C-stat/d.o.f.=441.1/314. The residuals from this fit showed prominent wavy structures centered around $\sim$8~keV, $\sim$14~keV, and $\sim$28~keV. We attempted first to improve the fit by adding a thermal component, as done in the case of \src,\ but the fit was insensitive to the addition of a hot blackbody with virtually no changes in the shape of the residuals. We tried changing the cut-off power-law with a more physically motivated comptonization model (as those introduced in Sect.~\ref{sec:src_spe}) but again we could not achieve improvements in the fit results. We found that the only way to improve the residuals around $\sim$8~keV and $\sim$14~keV was via the addition of an {\sc edge} and a {\sc gabs} component, respectively. This model provided a good description of the data (C-stat/d.o.f.=326.5/309). We measured an absorption column density of $N_{\rm H}$=(7.7$\pm$2.4)$\times$10$^{22}$~cm$^{-2}$, a power-law photon index of $\Gamma$=-1.3$\pm$0.4, a cut-off energy of $E_{\rm C}$=4.3$\pm$0.5~keV, and an energy (optical depth) of the edge of 6.99$\pm$0.08~keV (0.4$\pm$0.1). Concerning the {\sc gabs} component, from the fit we measured a centroid energy of 14.2$\pm$0.3~keV, a width of 1.9$\pm$0.4~keV, and a depth of 1.6$^{+0.7}_{-0.5}$~keV (translating into an optical  depth\footnote{\url{https://heasarc.gsfc.nasa.gov/xanadu/xspec/manual/node246.html}} of the {\sc gabs} component of 0.34$\pm$0.14). The estimated flux in the 2-40~keV energy band was (2.1$^{+0.1}_{-0.6}$)$\times$10$^{-11}$~$\ferg$. As the power-law photon index turned out negative, we also tested a possible improvement in the description of the apparently complex curvature of the source spectrum by employing a {\sc highecut*power-law} component in place of the simpler {\sc cutoffpl} component. In this case, we obtained an equivalently good description of the data  (C-stat/d.o.f.=312.8/308), measuring a power-law photon index of $\Gamma$=-0.3$^{+0.3}_{-0.4}$, a cut-off energy of $E_{\rm C}$=7.8$^{+0.4}_{-0.7}$~keV, and a folding energy of $E_{\rm F}$=5.7$^{+0.7}_{-0.6}$~keV. The best fit parameters for the edge and for the {\sc gabs} components remained virtually unchanged to within the associated 90\%~c.l. uncertainties. From the fit we obtained an energy (optical depth) of the edge of 6.99$\pm$0.08~keV (0.6$\pm$0.1), a {\sc gabs} centroid energy of 14.0$\pm$0.3~keV, a width of 1.6$\pm$0.3~keV, and a depth of 1.1$\pm$0.4~keV. The estimated optical depth of the {\sc gabs} component is thus 0.27$\pm$0.11. We also report that the normalization constants of the FPMA and FPMB with respect to the reference one fixed at 1.0 for XRT were both 0.43$\pm$0.07. This is expected due to the fact that the short XRT pointing of \srcb\ was carried out during a relatively bright portion of the \nustar\ observation (see Fig.~\ref{fig:nustar_lc2}). Inspecting the residuals from the fit with the last model illustrated above, hereafter the best fit model for \srcb,\ we still found the presence of moderately wavy structures around$\sim$28-30~keV. We thus tested the inclusion of an additional {\sc gabs} component around this energy. The improvement to the overall fit was modest (C-stat/d.o.f.=307.9/306), but interestingly the centroid energy of the second {\sc gabs} component ended up being compatible with double the value of the first component (25.7$\pm$1.9~keV). We measured for this component a depth of 1.0$_{-0.7}^{+0.9}$~keV, while the width was fixed to the same value of the first {\sc gabs} component as this parameter could not be constrained in the fit. None of the other model parameters were changed significantly by the introduction of the second {\sc gabs} component. We show the broad-band spectra of \srcb,\ together with the best fit model (not including the {\sc gabs} at $\sim$25~keV) and the residuals from the fit in Fig.~\ref{fig:bband2}.

To estimate the significance of the {\sc gabs} detection, we run a Monte Carlo Markov Chain (MCMC) using the Goodman-Weare sampling starting from the best-fit model (20 walkers, burning phase 6000, length 26\,000) for model with and without the gabs component. This is the same approach we employed in our previous paper, \citep{ferrigno20}. For simplicity, only the \nustar\ data were used in the simulation, as the XRT data could not contribute significantly to constrain the spectral parameters and the {\sc gabs} feature is outside the energy coverage of these data. We simulated 1000 spectra for model parameter sets extracted from the MCMC runs in each case. For each simulated spectrum, we performed a fit and recorded the value of the C-statistics obtained. For the model with the line, about 15\% of the simulated fits have a C-statistics larger than the one obtained from the data, showing a compatibility of the model with data. On the contrary, all values of the C-statistics obtained from simulations without the lines are well below the value obtained from the data. By extracting the average of the C-statistics obtained from the simulations and their standard deviations, we compute the difference with the value obtained from the data and estimate that the model without a {\sc gabs} component is rejected at 4.3$\sigma$ Gaussian-equivalent significance level\footnote{Note that, in presence of a sufficient number of channels, the central limit theorem can be applied and the C-statistic for the spectral fit can be assumed to be Gaussian distributed \citep{kaastra17}. For the present application, we verified the compliance to this assumption by plotting the histograms and the Gaussian distributions of the C-statistic values derived from the average and the standard deviation.}.

To consolidate the detection of the {\sc gabs} absorption feature, we also tested for completeness the applicability of more multi-component semi-phenomenological models, as suggested by \citet{doroshenko12} and \citet{tsy19}. We used a model comprising two comptonization components ({\sc compTT} in {\sc xspec}) with either the seed photon temperature linked to be the same or with free to vary temperatures. The model could not satisfactorily fit the data (C-stat/d.o.f.=455.1/309), with prominent residuals left at the energy of the edge and around $\sim$14~keV. Adding the edge slightly improved the fit (C-stat/d.o.f.=383.0/307) but the model is not able to definitively reproduce the shape of the absorption feature around $\sim$14~keV. We argue that this is likely related to the fact that the {\sc gabs} required by the broad-band spectra of \srcb\ is significantly narrower and deeper than the similar features fit with the double {\sc compTT} model in a few X-ray pulsars \citep{doroshenko12,tsy19}.

\begin{figure}
\hspace{-0.5cm}
  \includegraphics[width=6.5cm,height=8.9cm,angle=-90]{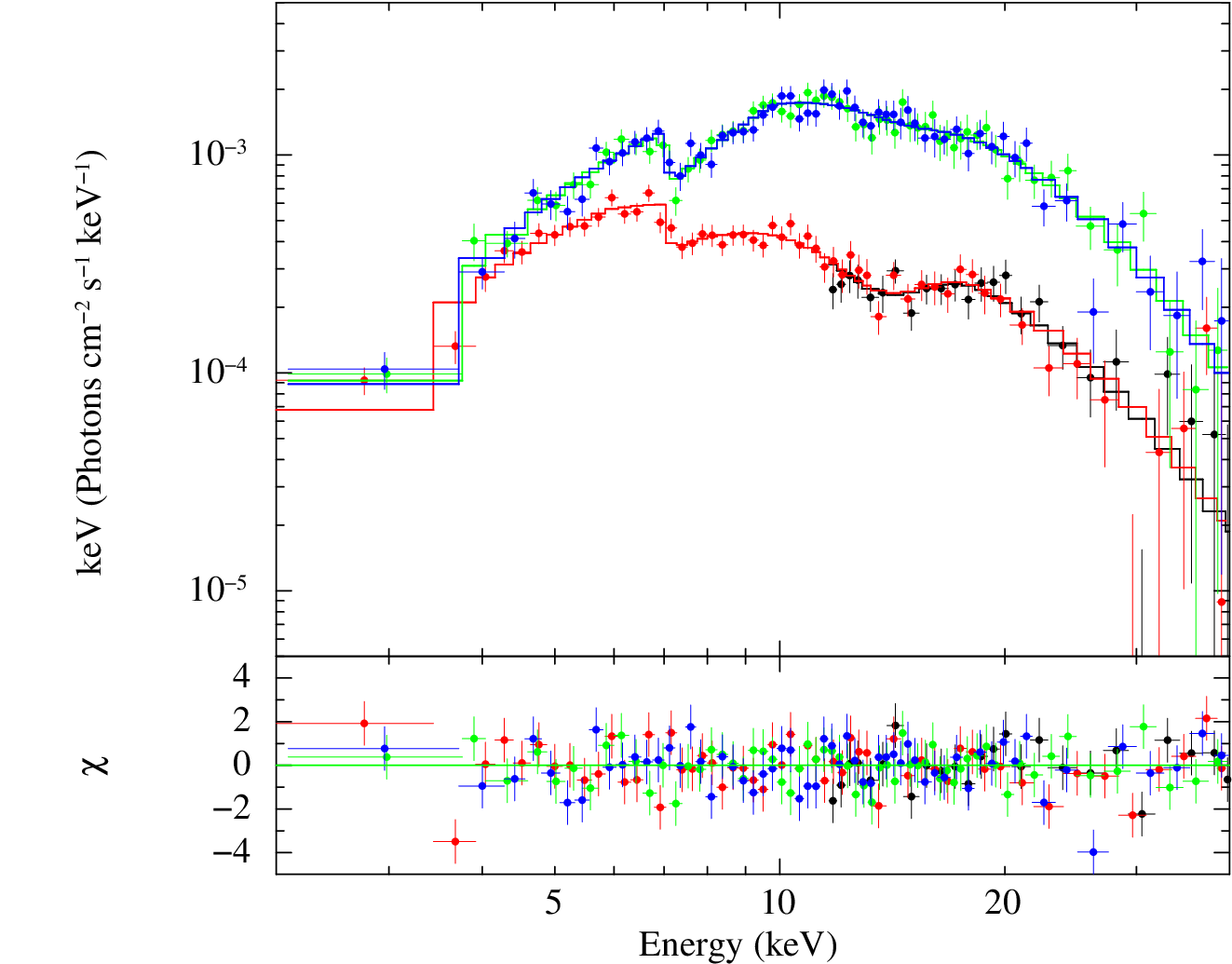}
  \caption{\label{fig:bband2hr} Unfolded \nustar\ HR-resolved spectra of \srcb.\ Spectra extracted during the time intervals corresponding to HR$<$2 are shown in black (FPMA) and red (FPMB). The spectra extracted during the time intervals corresponding to HR$>3$ are shown in green (FPMA) and blue (FPMB). The best fit model comprises an absorbed cut-off power-law ({\sc Tbabs*highecut*powerlaw} in {\sc xspec}), an absorption feature ({\sc gabs} in {\sc xspec}), and an {\sc edge} centered at $\sim$7~keV (see Sect.~\ref{sec:srcb_spe} for more details).
  The best fit model is shown in the upper panel of the figure with solid lines, while the residuals from the fit are reported in the bottom panel.}
\end{figure}

As anticipated in Sect.~\ref{sec:nustar}, we extracted two HR-resolved spectra from the \nustar\ observation of \srcb.\ The spectrum extracted during the time intervals when HR$>$3 had an effective exposure time of 7.04~ks, while the spectrum extracted during the time intervals corresponding to the case HR$<$2 has an effective exposure time of 24.8~ks. The average source count-rate measured by the FPMs during the HR$>$3 case was a factor of $\sim$3 larger (0.46$\pm$0.08) than that corresponding to the HR$<$2 case (0.15$\pm$0.03). This means that, taking into account the difference in the effective exposure time, the two sets of spectra were characterized by a similar number of counts. We fit simultaneously all spectra with the best fit model described above. The statistics was sufficiently good to avoid freezing any parameter in the fit, although for consistency we fixed in each of the two sets the parameters of the FPMB spectra to those of the FPMA spectra (leaving only the inter-calibration constant free to vary). We summarize the outcomes of the HR-resolved spectral analysis in Table~\ref{tab:fit}. The change in the HR is driven by both a modest variation of the absorption column density and a more substantial change in the continuum. No significant changes are observed in the parameters of the edge, while the {\sc gabs} absorption feature centered at $\sim$14~keV turned out substantially less deep in the case of the HR$>$3 spectra. Figure~\ref{fig:bband2hr} shows the four FPM spectra extracted at the different HRs and fit with the model summarized in Table~\ref{tab:fit} and  the residuals from the best fit.
\begin{table}
\caption{Best-fit model parameters obtained from the HR-resolved spectral analysis of the \nustar\ observation of \srcb.\ We report the parameters obtained from the fit with a model comprising an absorbed cut-off power-law component, including an edge and an absorption {\sc gabs} feature. The centroid energy (depth) of the edge is indicated with $E_\mathrm{edge}$ ($\tau_\mathrm{edge}$), while for the {\sc gabs} we indicated the centroid energy with $E_\mathrm{Cyc}$, the depth with $d_\mathrm{Cyc}$, the width with $\sigma_\mathrm{Cyc}$, and the optical depth as $\tau_\mathrm{Cyc}$. The flux corresponds to the power-law flux for the respective units and is identical for the two units. Uncertainties are at 1-$\sigma$ level}. \label{tab:fit}
\begin{tabular}{lr@{}lr@{}ll}
\hline
\hline
 & \multicolumn{2}{c}{HR $<$ 2} & \multicolumn{2}{c}{HR $>$ 3} &  \\
\hline
$N_\mathrm{H}$ &22 &$\pm$5 & 20 &$\pm$5 & $10^{22}$cm$^{-2}$\\
$\Gamma$ &0.9 &$_{-0.4}^{+0.3}$ & -0.15 &$\pm$0.20 & \\
$E_\mathrm{C}$ &10.8 &$\pm$0.4 & 9.8 &$\pm$0.4 & keV\\
$E_\mathrm{F}$ &7 &$_{-2}^{+2}$ & 6.2 &$\pm$0.5 & keV\\
$E_\mathrm{edge}$ &6.94 &$\pm$0.11 & 7.00 &$\pm$0.10 & keV\\
$ \tau_\mathrm{edge}$ &0.31 &$\pm$0.09 & 0.42 &$\pm$0.10 & \\
$E_\mathrm{Cyc}$ &13.1 &$\pm$0.2 & 13.15 &$_{-0.16}^{+0.22}$ & keV\\
$\sigma_\mathrm{Cyc}$ &2.7 &$_{-0.4}^{+0.5}$ & 0.2 &$\pm$0.3 & keV\\
$d_\mathrm{Cyc}$ &5 &$_{-2}^{+3}$ & 0.20 &$_{-0.08}^{+0.10}$ & keV \\
$\tau_\mathrm{Cyc}$ &0.74 &$\pm$0.38 & 0.4 &$\pm$0.6 & keV \\
Flux (3-20 keV) &22 &$_{-3}^{+7}$ & 75 &$_{-7}^{+9}$ & $10^{-12}\,\mathrm{erg\,s^{-1}\,cm^{-2}}$\\
\hline
\end{tabular}
\end{table}

\section{Discussion and conclusions}
\label{sec:conclusions}

In this paper, we reported on archival observations performed simultaneously with \swift\ and \nustar\ toward the SFXTs \src\ and \srcb.\ The unique combination of sensitivity and wide energy coverage of the FPMA and FPMB on-board \nustar\ allowed us to perform the first detailed broad-band analysis of the X-ray emission from the two sources. The \nustar\ data of both sources were complemented with simultaneous short \swift/XRT pointing to extend, in principle, the energy coverage down to $\sim$0.3~keV. As illustrated in Sect.~\ref{sec:intro}, one of the main objectives of the broad-band spectral analysis is to look for spectral features, as the CRSF which can provide a measurement of the NS magnetic field strength and possibly discriminate among the proposed models to interpret the SFXT phenomenology.

Concerning \src,\ the observations presented here caught the source during a relatively low emission state. Although a number of moderately faint flares were recorded and well visible in the \nustar\ lightcurves, we were not able to find any evidence of HR variation that could have hinted to spectral variability. Therefore, a single spectrum was extracted for the two FPMs and fit together with the XRT spectrum. Due to the faintness of the source and the high extinction in the direction of the source, the XRT spectrum was endowed with a low statistics and virtually no signal below 2~keV. The source 2-40~keV emission could be satisfactorily described by using a phenomenological model which is commonly adopted for X-ray binaries with NS accretor. This model comprises a hot blackbody and a power-law component, both affected by an accretion column density well above $\sim$10$^{22}$~cm$^{-2}$. Although pulsations have never been firmly detected from \src\ during previous observations and no evidence of coherent modulations were found in the current \nustar\ data, the parameters we measured for the hot blackbody indicates that a small portion of the NS surface ($\sim$300~m in diameter) is likely heated-up by the on-going accretion. As it occurs in many other X-ray binaries, pulsations might have gone undetected so far due to the small misalignment between the NS magnetic and rotation axes, or due to the slow rotation which can only be detected with much longer observations that have been performed to date \citep[note that the longest spin period measured so far in the case of a X-ray binary with a massive companion, possibly a supergiant star, is 36.2~ks; see][and references therein]{sidoli17}. No evidence for the presence of CRSFs could be found in the \nustar\ spectra of \src.\ We note that the previously reported feature at $\sim$15~keV marginally discussed by \citet{2012PASJ...64...99N} is centered on an energy which was not covered by the XIS and HXD/PIN on-board \suzaku.\ We thus argue that the feature might have most likely risen due to inter-calibration uncertainties between the two instruments in a delicate energy range of the source spectrum where the black-body and power-law components, as unveiled by \nustar,\ match one another to shape the source continuum (see Fig.~\ref{fig:bband1}).

The combined XRT and FPM spectra of \srcb\ proved to be  particularly interesting since no previous high-sensitivity broad-band study of the source emission had been presented in the literature. As in the case of \src,\ the XRT spectrum was endowed with limited statistics mainly due to the high extinction in the direction of the source (approaching 10$^{23}$~cm$^{-2}$, see Sect.~\ref{sec:results}) and could not help constraining the spectral parameters below $\sim$2~keV. The FPM spectra displayed a rather complex curvature which could be satisfactorily fit by using a cut-off power-law with the addition of an edge at $\sim$7~keV and the presence of a broad absorption feature centered around $\sim$14~keV, which statistical significance was assessed at 4.3~$\sigma$ (see Sect.~\ref{sec:results}). The centroid energy of the edge measured from the \nustar\ spectra is compatible with being due to neutral iron and it is typically observed in sources, including X-ray binaries, where some non-ionized material is intervening along the line of sight to the observer and (partially) hide the X-ray source \citep[see, e.g.,][and references therein]{koljonen18}. Edges have been observed in many highly obscured high mass X-ray binaries, HMXBs, including several SgXBs \citep[see, e.g.][and references therein]{walter2015}. These edges are often accompanied by prominent iron emission lines which are the effect of fluorescence from the intervening material along the line of sight which is also illuminated by the X-rays from the accreting compact object \citep[see, e.g., the case of the wind-fed highly obscured high mass X-ray binary IGR\,J16318-4848;][]{2007A&A...465..501I}. However, there have been cases where the fluorescence emission is not observed and only edges are present in the X-ray spectra of some X-ray binaries. In these cases, the intervening material is generally termed as a ``screen'' and it is necessarily assumed to be located far from the compact object to avoid direct illumination. Dipping sources are known to display edges without the presence of any fluorescence line during short obscuration episodes associated with ``dips'' in their lightcurves and broadly ascribed to the presence of structures at the accretion disk rim \citep[see, e.g., the case of AX J1745.6–2901;][]{hyodo08,ponti15,ponti18}. In high mass X-ray binaries, the presence of ``screens'' is less common, but at least in the case of the Be X-ray binary V\,0332+53, \citet{bykov21} reported the disappearance of the fluorescence lines during part of the 2004-2005 source outburst monitored with the \rxte/PCA, while the absorption edge at $\sim$7.1~keV was still clearly detected in the PCA spectra. The authors commented that such screens in the case of HMXBs could be related to the presence of large dense structures surrounding (but not located close to) the compact object, like wind or accretion stream through the binary system inner Lagrange point. Given the lack of a monitoring over time of the broad-band spectrum of \srcb,\ it is presently not possible to investigate further this possibility, as one would like to check how eventually the intervening screens affect the X-ray emission from the source at different orbital phases to eventually constrain the geometry and location of this material. The detection and evolution of screens over time in \srcb\ could be helpful in gaining further insights in the future on the accretion processes in this source, as well as in SFXTs as a class, as they could reveal the presence of structures connected with accretion streams and/or temporarily forming accretion disks, as already proposed in the case of some of these objects \citep{romano15}.

Of particular interest is the absorption feature at $\sim$14~keV detected in the broad-band spectrum of \srcb.\ In Sect.~\ref{sec:srcb_spe}, we tested at the best of our current possibilities the presence of this feature by verifying that alternative models could not reproduce the source spectral energy distribution as measured simultaneously by \swift\ and \nustar.\ If interpreted in terms of a CRSF, the absorption feature at $\sim$14~keV would indicate a relatively weak magnetic field for the NS hosted in \srcb,\ i.e. $\sim$1.2$\times$10$^{12}$~G \citep{staubert19}. The parameters measured for the {\sc gabs} component, and in particular the optical depth (reaching a value of $\sim$0.7 in the HR$<$2 spectra, see Table~\ref{tab:fit}), are comparable to those of similar features measured in several other HMXBs containing strongly magnetized pulsars and observed with \nustar.\ Noticeable cases are those of Her\,X-1 \citep{fuerst13}, A\,0535+26 \citep{ballhausen17}, RX J0520.5-6932 \citep{Tendulkar14}, and Cen\,X-3 \citep{tomar21}, where typical values of the {\sc gabs} optical depth range from 0.4 to $\gtrsim$1. A particularly striking case is that of the ultra luminous X-ray source  NGC 300 ULX, whose X-ray emission is characterized by a CRSF centered at $\sim$13~keV with an optical depth of $\sim$0.45 \citep{walton18}.  To further strengthen the CRSF interpretation, we noted in Sect.~\ref{sec:srcb_spe} possible residuals in the fit from the best determined model around an energy that is roughly double compared to the centroid of the CRSF, thus mimicking the presence of a first harmonic (although the statistics was not sufficient to draw a firm conclusion). The HR-spectral analysis revealed that the parameters of the absorption feature, as well as those of the continuum emission, changed significantly during the time intervals when HR$>3$ and HR$<$2. The mechanism driving these changes is hard to be understood in terms of the present data-sets, but we note that CRSFs are well known to change in energy, width, and depth when the spectral energy distribution of the source varies along flaring and outburst episodes \citep[see, e.g.][and references therein]{fuerst14,Vybornov17,fuerst18}.

As \srcb\ is one of the confirmed SFXTs with one of the highest measured dynamic range in the X-ray domain, this finding has deep implications on the applicability of the proposed models to interpret the behavior of these sources as a class. With a magnetic field as low as $\sim$1.2$\times$10$^{12}$~G, the possibility of halting accretion on the compact object in this system via the so-called magnetic gating would hardly be applicable \citep{bozzo08}. This means that only the centrifugal gating, associated to the well known propeller effect \citep{illarionov75,davies79,davies81}, or the settling accretion regime could be at work to limit accretion and lead to the low average luminosity usually observed from this source (see Sect.~\ref{sec:srcb}). However, as illustrated extensively by \citet{bozzo08} and \citet{bozzo17}, the centrifugal gating alone cannot be exploited to achieve dynamic ranges in the X-ray luminosity that exceed a factor $\sim$100 even in presence of a clumpy stellar wind \citep[see, also the discussions in][]{nunez17}. A similar conclusion applies in the case of the settling accretion regime, which typical reduction on the mass accretion rate compared to the undisturbed spherical case (or ``direct accretion'') is of a factor of $\sim$30-100 \citep{shakura12,postnov14}.

As the hypothesis that stellar winds in SFXTs are extremely clumpy or much weaker than those of supergiant stars in classical systems is currently not supported by the observations (Sect.~\ref{sec:intro}), a remaining possibility in the case of \srcb\ is that a substantial fraction of the X-ray dynamic range is due to the combination of the centrifugal gating/settling regime, operating in a clumpy wind environment, with the elongated and eccentric orbit. Although not firmly measured, the eccentricity in the case of \srcb\ has been estimated to be about 0.3-0.4, and the lowest emission state of the source unveiled thanks to upper limits on the source non-detection obtained with \xmm\ observations were demonstrated to be close to apastron (see Sect.~\ref{sec:srcb}). It is well known that in a wind-fed binary with a NS accretor, such elongated orbit can already produce a modulation of the mass accretion rate onto the compact object, and thus of the resulting X-ray luminosity, by a factor of $\sim$10-100 even in the context of a homogeneous and spherically symmetric wind \citep[taking into account also the effect of photoionization of the stellar wind by the NS X-ray radiation; see][and references therein]{bozzo21}. Exploring the applicability of all theoretical models to the case of SFXTs with relatively low magnetic fields is beyond the scope of the current paper, but we propose here that the total dynamic range in the X-ray luminosity measured so far from \srcb\ could effectively be the result of the convolution between a centrifugal inhibition/settling accretion regime and the effect of a possibly peculiar elongated orbit among the other known SFXT members. Compatibly with this scenario, we note that the total dynamic range recorded by \swift\/XRT during the monitoring campaign initiated after the onset of a source outburst in 2009 achieved a factor of $\sim$3000 only when considering also the upper limits on the source non-detection about $\sim$11~days from the peak of the event (i.e. when the NS was already moving close to apastron). This is a much less extreme variability than that displayed by the SFXT IGR\,J17544-2619, which achieved a dynamic range in the X-ray luminosity of $\sim$10$^{6}$ and is characterized by one of the shortest known orbital period among the SFXTs \citep[thus unlikely to have any eccentricity at all; see][]{romano15}. This conclusion might as well corroborate the idea discussed in several literature papers that the SFXTs constitute, after-all, an heterogeneous class of objects \citep{walter2015}. Extended and high sensitivity observations of \srcb\ endowed with a broad-band energy coverage along different parts of the system orbit (e.g. exploiting the combination of \xmm\ and \nustar) would be an invaluable asset to verify the proposed scenario and eventually look for pulsations from this system that so far remained elusive.

\section*{Data availability}
All data exploited in this paper are publicly available from the \nustar\ and \swift\  archives and processed with publicly available software.

\section*{Acknowledgements}
We thank the anonymous referee for useful and constructive comments.
EB and PR acknowledges financial contribution from contract ASI-INAF I/037/12/0.
This work made use of data supplied by the UK Swift Science Data Centre at the
University of Leicester \citep[see][]{2007A&A...469..379E,2009MNRAS.397.1177E}.

\bibliography{bib}{}
\bibliographystyle{mnras}

\appendix

\end{document}